\documentclass[preprint2]{aastex6}

\usepackage{ulem}
\usepackage{xcolor}
\usepackage{amsmath}
\usepackage{cases}
\usepackage[T1]{fontenc}

\begin{document}
\title{A unified accreting magnetar model for long-duration gamma-ray bursts and some stripped-envelope supernovae} 
\author{W. L. Lin\altaffilmark{1}, X. F. Wang\altaffilmark{1,2,3}, L. J. Wang\altaffilmark{4}, Z. G. Dai\altaffilmark{5,6}}
\altaffiltext{1}{Physics Department and Tsinghua Center for Astrophysics (THCA), Tsinghua University, Beijing 100084, China; \\
linwl@mail.tsinghua.edu.cn; wang\_xf@mail.tsinghua.edu.cn}
\altaffiltext{2}{Beijing Planetarium, Beijing Academy of Science and Technology, Beijing 100044, China}
\altaffiltext{3}{Purple Mountain Observatory, Chinese Academy of Science, Nanjing 210008, China}
\altaffiltext{4}{Astroparticle Physics, Institute of High Energy Physics, Chinese Academy of Sciences, Beijing 100049, China}
\altaffiltext{5}{School of Astronomy and Space Science, Nanjing University, Nanjing 210023, China}
\altaffiltext{6}{Key Laboratory of Modern Astronomy and Astrophysics (Nanjing University), Ministry of Education}

\begin{abstract}

Both the long-duration gamma-ray bursts (LGRBs) and the Type I superluminous supernovae (SLSNe~I) have been proposed to be primarily powered by central magnetars. A correlation, proposed between the initial spin period ($P_0$) and the surface magnetic field ($B$) of the magnetars powering the X-ray plateaus in LGRB afterglows, indicates a possibility that the magnetars have reached an equilibrium spin period due to the fallback accretion. The corresponding accretion rates are inferred as $\dot{M}\approx10^{-4}-10^{-1}$ M$_\odot$ s$^{-1}$, and this result holds for the cases of both isotropic and collimated magnetar wind. For the SLSNe~I and a fraction of engine-powered normal type Ic supernovae (SNe~Ic) and broad-lined subclass (SNe~Ic-BL), the magnetars could also reach an accretion-induced spin equilibrium, but the corresponding $B-P_0$ distribution suggests a different accretion rate range, i.e., $\dot{M}\approx 10^{-7}-10^{-3}$ M$_\odot$ s$^{-1}$. Considering the effect of fallback accretion, magnetars with relatively weak fields are responsible for the SLSNe~I, while those with stronger magnetic fields could lead to SNe~Ic/Ic-BL. Some SLSNe~I in our sample could arise from compact progenitor stars, while others that require longer-term accretion may originate from the progenitor stars with more extended envelopes or circumstellar medium. 

\end{abstract}


\section{Introduction}
\label{Sec:A: Introduction}

Rapidly rotating magnetars are promising central engine candidates for some transient astrophysical phenomena. They compete with black holes (BHs) to be the central engines of the gamma-ray bursts (GRBs; e.g., \citealp{1992Natur.357..472U, 1998A&A...333L..87D, 2001ApJ...552L..35Z}). They have been also proposed to power the X-ray flares and plateaus (shallow decays) in some GRB afterglows (e.g., \citealp{2006Sci...311.1127D, 2006ApJ...642..354Z, 2010MNRAS.409..531R, 2011A&A...526A.121D, 2013MNRAS.430.1061R, 2014ApJ...785...74L, 2018ApJS..236...26L, 2018ApJ...855...67L, 2018ApJ...869..155S}). In addition, the magnetar-powered model is invoked to interpret the luminosity evolution of different subclasses of supernovae (SNe; e.g., \citealp{2010ApJ...717..245K, 2010ApJ...719L.204W}), such as the normal SNe~Ic (e.g., \citealp{2018AA...609A.106T, 2019AA...621A..64T}), broad-lined SNe~Ic (SNe~Ic-BL; e.g., \citealp{2017ApJ...851...54W, 2017ApJ...837..128W}), and Type I superluminous supernovae (SLSNe~I; e.g., \citealp{2013ApJ...770..128I, 2017ApJ...842...26L, 2017ApJ...850...55N, 2017ApJ...840...12Y, 2018ApJ...865....9B, 2018ApJ...869..166V, 2019ApJ...872...90B, 2020MNRAS.497..318L}).

In those models, magnetars are usually treated as isolated neutron stars (NSs) which spin down due to the magnetic dipole radiation. In the context of core-collapse explosion, however, the stellar debris could circulate into a disk and interact with a nascent magnetar, which has a strong influence on the spin evolution and outflows of the magnetar (magnetic propeller; e.g., \citealp{2011ApJ...736..108P, 2018ApJ...857...95M}). For a given accretion rate, the magnetar could reach an equilibrium spin period, i.e., $P_\mathrm{eq}\propto B^{6/7}$, where $B$ is the surface magnetic field of the magnetar. Such a model has been invoked and further developed to study the diverse X-ray light curves of long- and short-duration GRB (LGRB and SGRB) afterglows \citep{2012ApJ...759...58D, 2014MNRAS.438..240G, 2017MNRAS.470.4925G, 2018MNRAS.478.4323G}, since magnetar-disk system could be formed in the cases of both core-collapse explosion and binary compact star mergers. Assuming that the magnetar wind is collimated, \citet{2018ApJ...869..155S} find a correlation between the surface magnetic field ($B$) and the initial spin period ($P_0$) of isolated magnetar engines for X-ray plateaus in GRB afterglows, in agreement with the $B\propto P_\mathrm{eq}^{7/6}$ relation for the accreting magnetars.

Based on the magnetic propeller model, we tentatively explore the properties of a portion of LGRBs and SNe that can be explained by considering a magnetar as a dominant power source. Our paper is organized as follows. In Section~\ref{Sec: sample}, we collect a sample of transients that are potentially powered by magnetars, including LGRBs with X-ray plateaus, SLSNe~I, SNe~Ic and SNe~Ic-BL. In Section~\ref{Sec: BPdistribution}, we show the $B-P_0$ distribution inferred from isolated magnetars as the central engines for different types of transients, and further discuss its physical implications. A summary is given in Section~\ref{Sec: conclusions}.

\section{Data and sample selection}
\label{Sec: sample}

\citet{2018ApJS..236...26L} systematically studied GRB X-ray plateaus, which are selected from the Neil Gehrels $Swift$/XRT data observed during 2004 December $-$ 2017 May, and concluded that 19 LGRB X-ray plateaus could be explained by the energy injection from isotropic magnetar wind. Assuming that the magnetar wind is collimated in the plateau phase, however, \citet{2014ApJ...785...74L} found more potentially magnetar-powered events from the XRT data obtained between 2005 January and 2013 August. Note that four X-ray plateaus in LGRB  afterglows (GRB 060526, GRB 061110A, GRB 070110 and GRB 120422A) can be explained in both scenarios. We include both of the above two magnetar candidate samples in the following analysis, since the wind configuration is still debated. 

\begin{table*}
 \centering
  \caption{The samples, models and references \label{Tab: sample}}
  \setlength{\tabcolsep}{2mm}{
  \begin{tabular}{cccccc}
  \hline
  Transients  &  Number  &  Power source  & References \\
  \hline
  LGRB X-ray plateaus   &    19  &  Magnetar (isotropic wind)   & \citet{2018ApJS..236...26L}   \\
  LGRB X-ray plateaus   &   36   &  Magnetar  (collimated wind)  &     \citet{2014ApJ...785...74L} \\
  SLSNe~I   &   61   &   Magnetar   &  \citet{2017ApJ...850...55N, 2018ApJ...865....9B}; \\
                    &         &                     &   \citet{2018ApJ...869..166V, 2019ApJ...872...90B};\\
                   &         &                     &   \citet{2020MNRAS.497..318L}    \\
  SNe~Ic-BL without detected LGRBs   &    11  &  Magnetar/Magnetar+$^{56}$Ni    &     \citet{2017ApJ...851...54W} \\
  LGRB-SNe   &  2   &   Magnetar/Magnetar+$^{56}$Ni     &  \citet{2015Natur.523..189G, 2017ApJ...837..128W}  \\
  SNe~Ic                                 &  2   &   Magnetar+$^{56}$Ni     &  \citet{2018AA...609A.106T, 2019AA...621A..64T} \\
   \hline
\end{tabular}}
\end{table*}

The sample of SLSNe~I are mainly collected from \citet{2017ApJ...850...55N} and \citet{2018ApJ...869..166V}, which analyzed multi-band light curves of a total of 58 spectroscopically identified SLSNe~I based on magnetar-powered model. In addition, three more SLSNe I (PS16aqv, SN 2017dwh and SN 2018hti; \citealp{2018ApJ...865....9B, 2019ApJ...872...90B, 2020MNRAS.497..318L}) are also included in our sample. 

\citet{2017ApJ...851...54W} invoked magnetar as an alternative energy source to model the light curves and velocity evolution of 11 SNe~Ic-BL without detections of companion LGRBs. For SN 2007ru, SN 2010ah and PTF10qts, we collect $B$ and $P_0$ inferred from the pure-magnetar model, while the parameters for the other 8 events are determined based on the fits with the magnetar plus $^{56}$Ni model, which provides better fits with lower $\chi^2/$d.o.f. values. 

Magnetar model is also proposed to account for the high luminosity of SN 2011kl \citep{2015Natur.523..189G,2017ApJ...850..148W}, which is associated with ultra-long GRB 111209A. \citet{2017ApJ...837..128W} fitted the bolometric light curve of SN 1998bw (associated with GRB 980425) with magnetar plus $^{56}$Ni model, and found that the peak and tail of the light curve can be explained by magnetar spin-down. In addition to these two SNe associated with LGRBs (LGRB-SNe), two normal SNe~Ic (iPTF15dtg, \citealp{2019AA...621A..64T}; PTF11mnb, \citealp{2018AA...609A.106T}) that are also likely powered by magnetars are included in our sample. 

The information of our sample is tabulated in Table~\ref{Tab: sample}. From the references listed in Table~\ref{Tab: sample}, we collect the parameters ($B$ and $P_0$) inferred from the models that invoke a magnetar as the dominant energy source.

\section{$B-P$ distribution}
\label{Sec: BPdistribution}

\subsection{X-ray plateaus in LGRB afterglows}
\label{subsec: LGRBs}
\begin{figure*}[htbp!]
\center
\includegraphics[angle=0,width=0.5\textwidth]{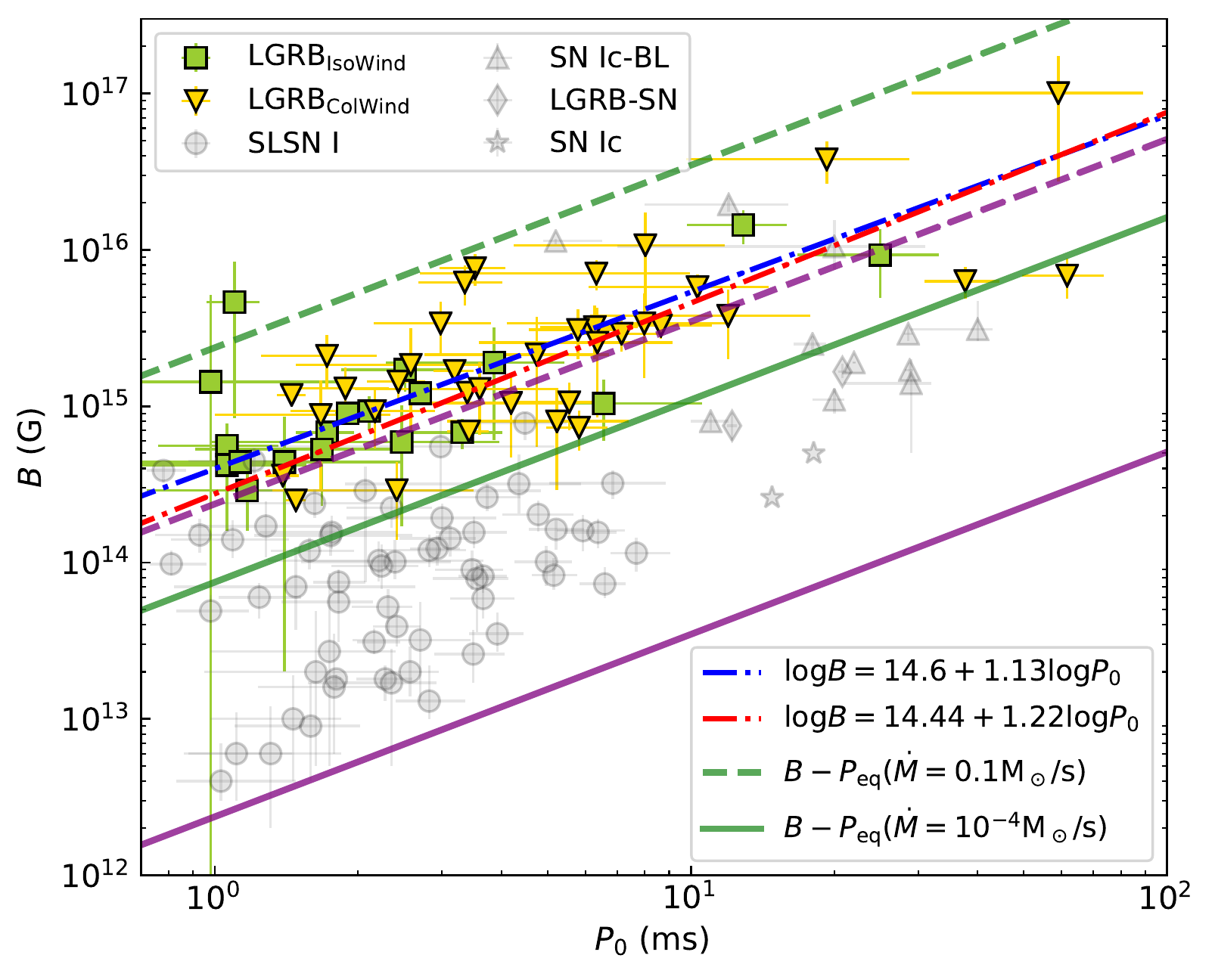}\includegraphics[angle=0,width=0.5\textwidth]{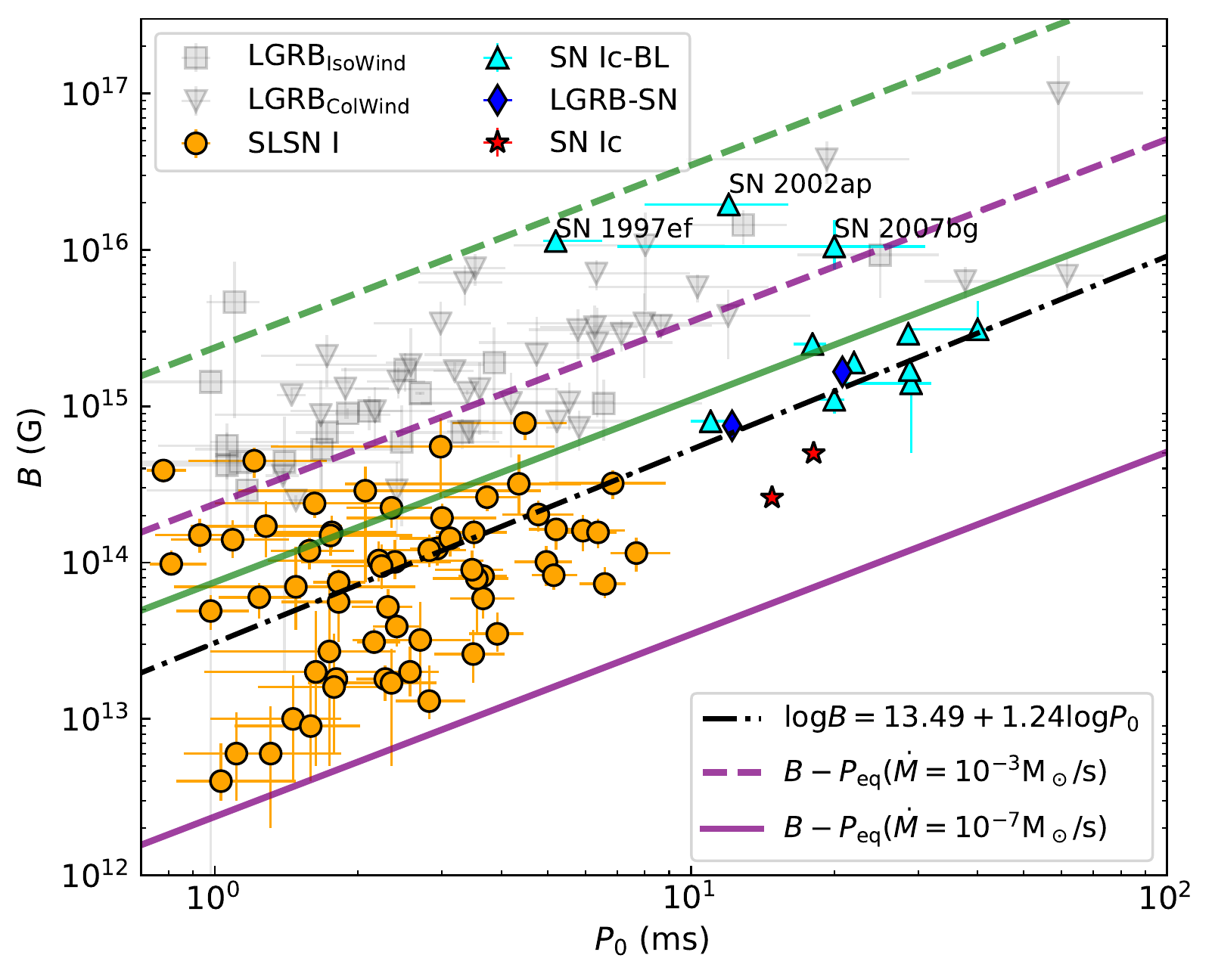}
\caption{Left panel: $B-P_0$ distribution for the magnetar engines of LGRB X-ray plateaus (green squares for the isotropic wind model; yellow triangles for the collimated wind model), SLSNe~I (gray circles), SNe~Ic-BL without LGRBs (gray triangles), LGRB-SNe (gray diamonds) and the normal SNe~Ic (gray stars). We show the fitting results for LGRB X-ray plateaus (blue dashed-dotted line for the isotropic wind model represented by Equation~\ref{eq: LGRB_iso}; red dashed-dotted line for the collimated wind model shown by Equation~\ref{eq: LGRB_col}). The expected $B-P_\mathrm{eq}$ correlations (Equation~\ref{eq: Peq}) are displayed assuming accretion rates $\dot{M}=0.1$ (green dashed line), $10^{-3}$ (purple dashed line), $10^{-4}$ (green solid line) and $10^{-7}$ (purple solid line) M$_\odot$ s$^{-1}$, respectively. Right panel: We highlight the $B-P_0$ distribution for SLSNe~I (orange circles), SNe~Ic-BL without LGRBs (cyan triangles), LGRB-SNe (blue diamonds) and the normal SNe~Ic (red stars). The black dashed-dotted line represents the best-fit correlation for these SNe (Equation~\ref{Eq: BP_SNe}). SN 1997ef, SN 2002ap and SN 2007bg are labeled due to their significant deviations from the fitting result. The LGRB X-ray plateaus are marked in gray symbols. Data references are given in Table~\ref{Tab: sample}.}
\label{fig: PB}
\end{figure*}

\begin{figure}[htbp!]
\center
\includegraphics[angle=0,width=0.46\textwidth]{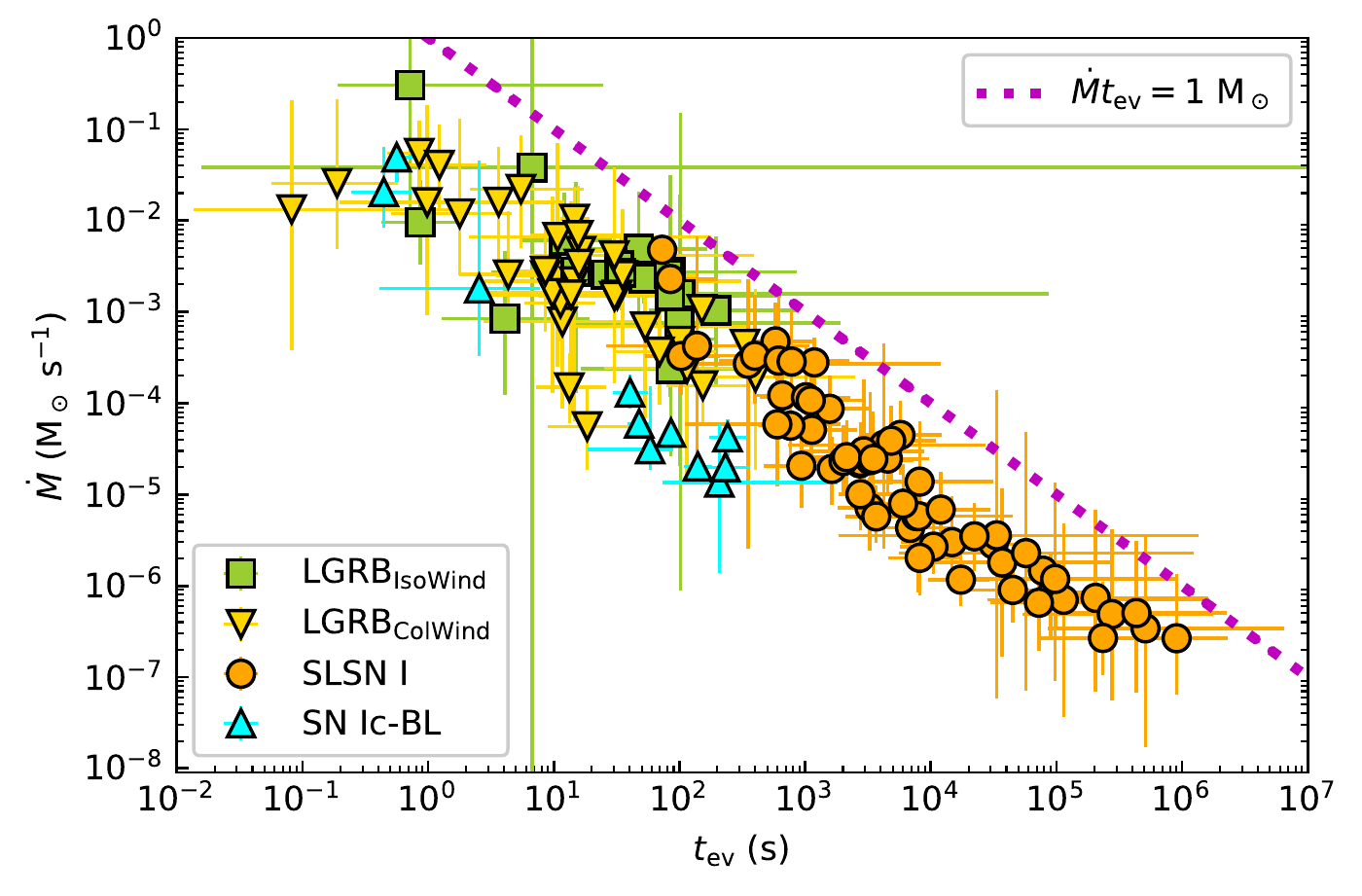}
\caption{The accretion rate ($\dot{M}$; Equation~\ref{eq: Peq}) versus the lower limit of accretion timescale ($t_\mathrm{ev}$; Equation~\ref{eq: Tev}) inferred for the magnetar engines of LGRB X-ray plateaus (green squares for the isotropic wind model; yellow triangles for the collimated wind model), SLSNe~I (orange circles) and SNe~Ic-BL without detections of LGRBs (cyan triangles).}
\label{fig: MdotT}
\end{figure}

The X-ray plateaus in some LGRB afterglows are observed to persist for $\sim100-10^5$ s before the steeper decline. Assuming that the X-ray plateaus are powered by the isotropic wind from the magnetars, the light curves can be used to constrain the surface magnetic field and initial spin period of magnetars. As seen in Figure~\ref{fig: PB}, magnetars with $P_0\sim1$ ms usually possess $B\sim10^{14}-10^{15}$ G, while those with $P_0\gtrsim10$ ms are accompanied by strong magnetic field of $B\sim10^{15}-10^{16}$ G. We perform a linear fit (See Appendix~\ref{Sec: fit} for the detailed descriptions of the fitting) to the $\log B-\log P_0$ distribution, and find
\begin{equation}
\log B=14.6^{+0.04}_{-0.05}+(1.13^{+0.11}_{-0.09})\log P_0,
\label{eq: LGRB_iso}
\end{equation}
where $P_0$ is in unit of millisecond. Such a correlation is consistent with the spin equilibrium state for the accreting magnetars ($B\propto P_\mathrm{eq}^{7/6}$ for a given accretion rate; e.g., \citealp{2011ApJ...736..108P}; see also Equation~(\ref{eq: Peq})). It implies that the initial spin period inferred from observations ($P_0$) could deviate from that of the magnetar at birth, but possibly corresponds to the equilibrium spin period as a result of interaction between the magnetar and surrounding accretion disk. The accretion rates of the disks are inferred as $\dot{M}\approx10^{-4}-0.1$ M$_\odot$ s$^{-1}$. We further estimate the evolutionary timescales for these magnetars to reach the spin equilibrium ($t_\mathrm{ev}\propto B^{-8/7}\dot{M}^{-3/7}$; \citealp{2018ApJ...857...95M}; see also Equation~(\ref{eq: Tev})), which turn out to be $\sim0.1-1000$ s. We consider $t_\mathrm{ev}$ as the lower limits for the accretion timescales ($t_\mathrm{acc}$) and show them in Figure~\ref{fig: MdotT}. Assuming $M_\mathrm{d}\sim\dot{M}t_\mathrm{acc}$, the total mass of accretion disk can be constrained to be $M_\mathrm{d}\gtrsim10^{-3}-0.5$ M$_\odot$. We caution that, a magnetar could possibly collapse into a BH, if it accretes a significant amount of materials and exceeds the maximum mass of a stable NS\footnote{Some pulsars with mass of $\sim2$ M$_\odot$ have been discovered \citep{2010Natur.467.1081D, 2013Sci...340..448A}, which sets a lower limit for the maximum mass of NSs ($M_\mathrm{max}$). Hitherto, there is no consensus on the upper limit of $M_\mathrm{max}$ (e.g., \citealp{2014PhRvD..89d7302L, 2017ApJ...850L..19M}).}. Hence, we assume an accretion disk mass of $\lesssim1$ M$_\odot$. Such a disk mass corresponds to an accretion timescale of $\lesssim10-10^4$ s, in agreement with the fallback timescale derived for Wolf-Rayet (WR) stars, i.e. $\sim10^2-10^5$ s\footnote{The radii ($r_\mathrm{e}$) of the envelopes of WR stars are $\sim10^{10}-10^{12}$ cm \citep{1995A&A...299..503K}. The free-fall timescale of the extended envelopes can be estimated by $t_\mathrm{ff}\sim(r_\mathrm{e}^3/GM)^{1/2}$, i.e. $100\lesssim t_\mathrm{ff}\lesssim10^5$ s.}; for a shorter timescale, the fallback materials could come from the core of the progenitors, which suggests an origin of compact progenitor stars for some LGRBs (e.g., \citealp{2006Natur.442.1008C, 2006ApJ...637..914W}). 

Notice that only $\sim20\%$ X-ray plateaus out of \citet{2018ApJS..236...26L} sample (including LGRB and SGRB X-ray plateaus) are consistent with the energy budget of magnetars, if the magnetar wind is isotropic. They argue that the BHs could be the central engines for most of the X-ray plateaus. If fallback accretion plays a role in the evolution of magnetars, the accreting magnetars could maintain the spin equilibrium on a longer timescale in the presence of an accretion disk, which may provide a natural explanation for the energy that is beyond the millisecond magnetar budget in some cases. Actually, the configuration of magnetar wind is still debated in the context of LGRB afterglow. Magnetar wind could escape via a collimated jet shortly after the SN explosion \citep{2009MNRAS.396.2038B}; on a longer timescale, wind could still be channeled into the polar region where the preceding jet drill its way out of the stellar envelope. A large number of X-ray plateaus can be explained by the injection of collimated magnetar wind alongside the GRB jets \citep{2014ApJ...785...74L, 2018ApJ...869..155S}. In Figure~\ref{fig: PB}, we compare the $B-P_0$ distributions that are derived from different wind models. Although the broader distributions are suggested in the case of collimated wind, they also follow a similar $B-P_0$ correlation (see also Appendix~\ref{Sec: fit}), i.e. 
\begin{equation}
\log B=14.44^{+0.03}_{-0.03}+(1.22^{+0.07}_{-0.06})\log P_0.
\label{eq: LGRB_col}
\end{equation}
And the inferred mass inflow rates are similar to those obtained by \citet{2018ApJ...869..155S} and are also consistent with the results based on the isotropic wind model. Therefore, both results suggest that the central magnetars could experience interactions with the surrounding accretion disks and finally reach a spin equilibrium.

\subsection{SLSNe~I, SNe~Ic and SNe~Ic-BL}
\label{subsec: SLSNe}
Core-collapse SNe usually reach their peak luminosities at tens of days after explosion. Around that time, magnetar wind should be near-isotropic and contributes most of the rotation energy to heat and/or accelerate the ejecta.

Based on the model fits to the light curves \citep{2017ApJ...850...55N, 2018ApJ...865....9B, 2018ApJ...869..166V, 2019ApJ...872...90B, 2020MNRAS.497..318L}, the magnetar engines of SLSNe~I are characterized by magnetic field of $10^{12}<B\lesssim 10^{14}$ G and spin with a short period of $1\lesssim P_0<10$ ms (Figure~\ref{fig: PB}). The magnetars with longer initial spin period appear to have stronger magnetic field. This correlation possibly suggests that the engine timescale is roughly comparable to the diffusion timescale of ejecta, which corresponds to $B\propto P_0$ \citep{2017ApJ...850...55N}. If the physics of accretion-induced spin equilibrium could apply to the cases of SLSNe~I, the inferred accretion rates  based on Equation~(\ref{eq: Peq}) mainly fall between $10^{-7}-10^{-3}$ M$_\odot$ s$^{-1}$.

SNe~Ic-BL might be accompanied by the birth of magnetars, since an upper limit of the kinetic energy of SNe~Ic-BL (i.e. $\sim$ a few $10^{52}$ erg s$^{-1}$, model dependent though) is comparable to the maximum rotation energy of a millisecond magnetar \citep{2014MNRAS.443...67M}. Moreover, some SNe~Ic-BL (e.g., SN2010ay and PTF10vqv) are unlikely to be explained by radioactive $^{56}$Ni model and hence magnetar is invoked as a dominant power source \citep{2017ApJ...851...54W}. As seen from Figure~\ref{fig: PB}, most SNe~Ic-BL without coincident LGRBs in our sample invoke a magnetar with $P_0\gtrsim10$ ms and $B\sim10^{15}$ G, while some events require the magnetic field to be as strong as $\sim10^{16}$ G. Associated with GRB 980425, SN 1998bw requires $B\approx1.66\times10^{15}$ G for the central magnetar, in agreement with most of its non-GRB peers. Compared with the SLSNe~I magnetars, the magnetar candidates for SNe~Ic-BL have stronger magnetic field and longer initial spin period. 

\citet{2015Natur.523..189G} reported the observations of a luminous SN Ic SN 2011kl (associated with GRB 111209A), which has intermediate luminosity lying between typical SNe~Ic-BL and SLSNe~I. Following their fitting results, SN 2011kl could be also powered by a magnetar with an initial spin period of $P_0\approx12$ ms and magnetic field of $B\approx7.5\times10^{14}$ G. Two normal SNe~Ic (iPTF15dtg and PTF11mnb) require weaker magnetic field than most SNe~Ic-BL and LGRB-SNe in our sample \citep{2018AA...609A.106T, 2019AA...621A..64T}.

We note that, although SLSNe~I exhibit distinct spectral features at early times (e.g., \citealp{2011Natur.474..487Q, 2018ApJ...855....2Q}), their post-peak spectra eventually evolve to resemble those of SNe~Ic/Ic-BL \citep{2010ApJ...724L..16P, 2017ApJ...845...85L, 2019ApJ...872...90B, 2019ApJ...871..102N}. This suggests an underlying connection among these subtypes of SNe. By fitting the $B-P_0$ distribution of our SNe samples (including SLSNe~I, SNe~Ic and SNe~Ic-BL), the following correlation (see also the fitting  procedure described in Appendix~\ref{Sec: fit}) is derived
\begin{equation}
\log B=13.49^{+0.1}_{-0.1}+(1.24^{+0.14}_{-0.14})\log P_0,
\label{Eq: BP_SNe}
\end{equation}
which is consistent with $B-P_\mathrm{eq}$ correlation expected for the accreting magnetars at the spin equilibrium state. Although SNe~Ic/Ic-BL require different properties of magnetars than SLSNe~I, similar accretion rates are inferred for most SNe in our sample\footnote{Note that three SNe~Ic-BL (SN 1997ef, SN 2002ap and SN 2007bg) deviate significantly from the best-fit power-law relation for all SNe sample (Equation~\ref{Eq: BP_SNe}) in the $B-P_0$ diagram (Figure~\ref{fig: PB}). They might represent a subset of SNe~Ic-BL which are powered by magnetars with magnetic field being as strong as $\sim 10^{16}$ G. However, we caution that there could be some other reasons for the deviations. Due to the lack of stringent constraints on the magnetar or $^{56}$Ni contribution in fitting with magnetar plus $^{56}$Ni model, it is difficult to obtain the accurate parameters of the magnetar. In addition, diverse power sources might also lead to a deviation, since both of the magnetar plus $^{56}$Ni model and the two component pure-$^{56}$Ni model can provide viable explanation for the emission of SN 1997ef, SN 2002ap and SN 2007bg \citep{2003ApJ...593..931M, 2010A&A...512A..70Y}.}. Therefore, nascent magnetar plus the accretion disk system provide a unified picture to explain the production of SLSNe~I, SNe~Ic and SNe~Ic-BL. Based on the magnetic propeller model, the central magnetar with low magnetic field will be accelerated to a millisecond spin period, responsible for the energy source powering an SLSN~I; conversely, stronger magnetic field leads to longer equilibrium spin period of a magnetar, and it hence produces an SN~Ic/Ic-BL.

Assuming that magnetars can always reach the spin equilibrium state, the lower limits for accretion timescales in the cases of SLSNe~I are estimated as $t_\mathrm{ev}\sim10^2-10^6$ s based on Equation~(\ref{eq: Tev}), while SNe~Ic-BL show a similar $t_\mathrm{ev}$ distribution to LGRB X-ray plateaus (Figure~\ref{fig: MdotT}). Although the limits of accretion timescales for most SLSNe~I are consistent with the fallback timescale for the envelopes of the compact progenitor stars, a fraction of SLSNe~I require longer accretion timescales. Assuming that the accretion timescale ($t_\mathrm{acc}$) is equivalent to the fallback timescale, the long-term accretion could be due to the fallback of the stellar envelope from large radii or inner ejecta that cannot escape from the central object \citep{1989ApJ...346..847C, 2013ApJ...772...30D}. Since early-time bumps observed in some events could be attributed to the cooling of an shocked envelope with a radius of $\gtrsim 500 R_\odot$ (e.g., \citealp{2015ApJ...808L..51P, 2016MNRAS.457L..79N, 2016ApJ...818L...8S}), the immediate progenitors of a portion of SLSNe~I could be surrounded by largely extended envelopes, in agreement with the first possible scenario allowing for the late accretion. Alternatively, those bumps might suggest the possible existence of circumstellar medium (CSM; \citealp{2012A&A...541A.129L}). In this scenario, reversed shock could be produced by the ejecta-CSM interaction. Then inner layer of the ejecta would be decelerated by the reverse shock and finally bound to the gravity of the newborn magnetar, contributing to the late accretion. It remains unknown whether SLSNe~I could be associated with LGRBs. But it might be a challenge for an LGRB jet to push through the extended envelope or CSM surrounding the progenitor stars of some SLSNe~I. For this subclass of SLSNe~I, the magnetic field strength is found to be lower than $10^{14}$ G, which is inconsistent with the magnetic field required by LGRBs.

\begin{figure}[htbp]
\center
\includegraphics[angle=0,width=0.46\textwidth]{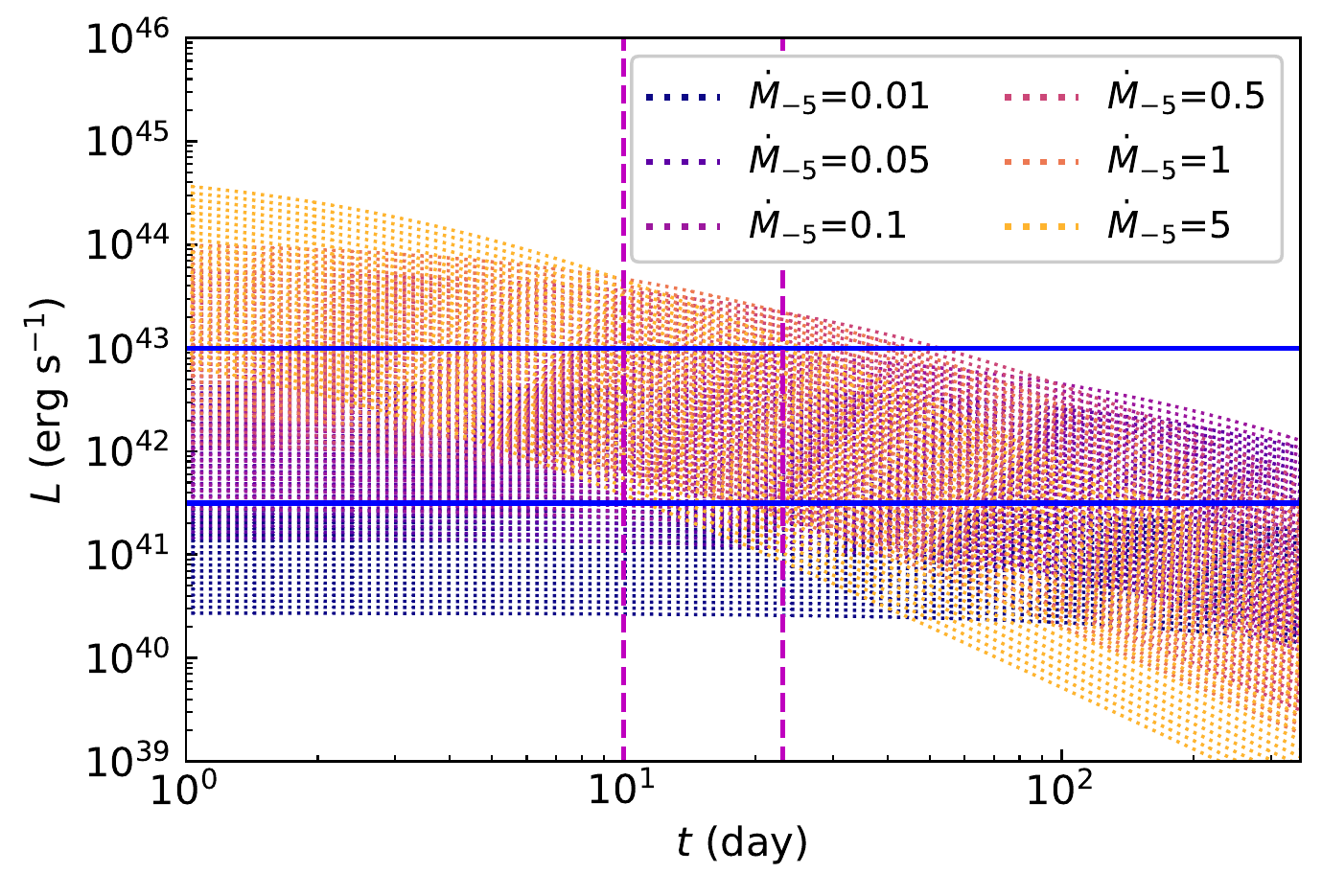}
\caption{The spin-down luminosity evolution of magnetars with $B$ and $P_0$, which are adopted (1) from the parameter space limited to $B\in[10^{14}, 10^{16}]$ G and $P_0\in [10, 100]$ ms, and (2) based on the $B-P$ relation for an accreting magnetar with the accretion rates $\dot{M}_{-5}=\dot{M}/(10^{-5}M_\odot/s)=0.01, 0.05, 0.1, 0.5, 1, 5$, respectively. The blue horizon lines show the peak luminosity range of the normal SNe~Ic ($41.5<\log L_\mathrm{p,Ic}<43$), and the purple vertical lines mark the major range of the peak time ($\approx10-23$ days; \citealp{2016MNRAS.458.2973P}).}
\label{fig: LT_Ic}
\end{figure}

Figure~\ref{fig: LT_Ic} shows some examples of the spin-down luminosity (magnetic dipole radiation luminosity) evolution of isolated magnetars with $B=10^{14}-10^{16}$ G and $P=10-100$ ms. The spin-down luminosities of those magnetars usually decline below $10^{44}$ erg s$^{-1}$ after 10 days since explosion, which are insufficient to power SLSNe~I according to the Arnett's rule (the peak luminosity equals the heating luminosity at peak time). Based on the peak luminosity statistics conducted by \citet{2016MNRAS.458.2973P}, the majority of normal SNe~Ic peak at a luminosity of log$(L_\mathrm{p,Ic})\approx41.5-43$ and at an epoch of $10-23$ days after explosion. Since the spin-down luminosities can match the peak luminosities of normal SN~Ic during their peak time, it is possible that those magnetars can produce emissions resembling the normal SNe~Ic. For an accretion rate between $10^{-7}-10^{-4}$ M$_\odot$ s$^{-1}$, the accreting magnetars with magnetic field of $\sim10^{14}-10^{15}$ G could reach an equilibrium spin period of $10-100$ ms, and hence possibly power normal SNe Ic. iPTF15dtg exhibits slowly declining light curves and strong O\textsc{i}~$\lambda$7774 emission at late phase, which can be explained in the magnetar-powered scenario \citep{2019AA...621A..64T, 2019ApJ...871..102N}. However, most SNe~Ic lack the engine-powered signatures, and $^{56}$Ni decay may be the power source. SLSNe~I and SNe~Ic-BL are preferentially found in dwarf galaxies with low metallicity (e.g., \citealp{2014ApJ...787..138L, 2016ApJ...830...13P, 2018MNRAS.473.1258S, 2020ApJ...892..153M}). With such low metallicity, stellar wind from massive progenitor stars might be reduced and hence the angular momentum can be sustained to help form a rapidly rotating magnetar. As most SNe~Ic prefer higher metallicity environments (e.g., \citealp{2020ApJ...892..153M}), it is thus expected that only a small subset of SNe~Ic are accompanied by the birth of magnetars.

\section{Discussions and conclusions}
\label{Sec: conclusions}
By involving the fallback accretion effect in the magnetar-powered scenario, we tentatively shed light on the correlation between the surface magnetic field ($B$) and initial spin period ($P_0$) of magnetar candidates that can account for the emissions of transients, such as LGRB X-ray plateaus, SLSNe~I, SNe~Ic and SNe~Ic-BL. 

In the context of the LGRB with X-ray plateaus, it is still debated whether the magnetar wind is isotropic and collimated in the plateau phase. No matter which wind configuration, however, the initial spin periods and the surface magnetic field of the magnetars are found to follow a positive correlation in agreement with $B\propto P_0^{7/6}$, suggesting that fallback accretion could play a role in the spin evolution of the magnetars. Thus, the initial spin period inferred from observations could correspond to the equilibrium spin period as a result of interaction between magnetar and surrounding accretion disk. Based on the magnetic propeller model, the accretion rates of the surrounding disks are inferred to be $\dot{M}\approx10^{-4}-10^{-1}$ M$_\odot$ s$^{-1}$ and the lower limits for the required accretion timescales are shorter than 1000 s, which are comparable to or shorter than the fallback timescale of the envelope of a WR star. This result is consistent with the proposal of compact progenitor stars as the progenitors of LGRBs, assuming that disk materials fall back from stellar envelope.

Magnetars are also invoked as an alternative engine to power SNe. On a timescale comparable to the rise time of core-collapse SNe, the magnetar wind should be isotropic. For SLSNe~I, SNe~Ic and SNe~Ic-BL in our magnetar-powered sample, $B-P_0$ distribution is found to be also consistent with the physics of accretion-induced spin equilibrium. The inferred accretion rates can be as low as $\dot{M}\approx10^{-7}-10^{-3}$ M$_\odot$ s$^{-1}$. Based on the magnetic propeller model, the central magnetars with low magnetic fields will be accelerated to a millisecond spin period, and may be able to power SLSNe~I; conversely, those with stronger magnetic fields could result in longer equilibrium spin period, which could explain a portion of SNe~Ic/Ic-BL.

For SLSNe~I, the accretion timescales are required to be $t_\mathrm{acc}\gtrsim10^2-10^6$ s. Assuming that accretion timescale is approximately equal to the fallback time, some SLSNe~I of our sample could be produced in the explosions of typical WR stars, while the others that require longer-term accretion could originate from the progenitor stars surrounded by more extended envelopes or CSM, which reconciles with the implication of the early bumps observed in some events (e.g., \citealp{2012A&A...541A.129L, 2015ApJ...808L..51P, 2016MNRAS.457L..79N, 2016ApJ...818L...8S}).

Above discussions are based on a simple method, with which we estimate the constant accretion rates from $B-P_0$ distribution inferred from the magnetar-powered models. However, besides the early-time accretion with constant rate, the late-time decline of accretion rate can also influence the spin period of the magnetar and hence magnetar outflow can deviate from the magnetic dipole radiation luminosity evolution (e.g., \citealp{2011ApJ...736..108P, 2018ApJ...857...95M}). We expect that more observations and detailed modeling work in the future will provide further clues for the nature of these GRBs and stripped-envelope SNe.

\section*{Acknowledgements}

We thank the anonymous referee for his/her constructive comments which help improve the manuscript. This work is supported by the National Natural Science Foundation of China (NSFC grants 12033003, 11633002, 11761141001, and 11833003) and the National Program on Key Research and Development Project (grant 2016YFA0400803 and 2017YFA0402600). L.J.W. acknowledges support from the National Program on Key Research and Development Project of China (grant 2016YFA0400801).

%\clearpage

{}
%\begin{appendix}
\appendix
\section{magnetic propeller model}
\label{Sec: scenario}

The interaction between a magnetar with its surrounding disk has an effect on the spin evolution and hence the outflows of the magnetar (e.g., \citealp{2011ApJ...736..108P, 2014MNRAS.438..240G, 2017MNRAS.470.4925G, 2018MNRAS.478.4323G, 2018ApJ...857...95M}). The interaction process depends on the relative locations of Alfv\'en radius ($r_\mathrm{m}$), co-rotation radius ($r_\mathrm{c}$) and light cylinder radius ($r_\mathrm{L}$). Alfv\'en radius is usually considered as the inner radius of the disk, where the ram pressure of the inflowing materials balances with the magnetic pressure of the magnetar. It can be given by 
\begin{equation}
r_\mathrm{m}=(GM)^{-1/7}R^{12/7}B^{4/7}\dot{M}^{-2/7},
\end{equation}
where $G$ is the gravitational constant, $M$/$R$/$B$ denotes the mass/radius/magnetic field strength of the central magnetar, and $\dot{M}$ is the mass inflow rate at the inner edge of the disk. Given that inflowing materials rotate at the local Keplerian angular velocity, i.e. $\Omega_\mathrm{K}=(GM/r^3)^{1/2}$, their co-rotation with the magnetar, occurs at a radius of 
\begin{equation}
r_\mathrm{c}=(GM/\Omega^2)^{1/3},
\end{equation}
where $\Omega=2\pi/P$ and $P$ are the angular velocity and spin period of the magnetar, respectively. The radius of the light cylinder is defined as $r_\mathrm{L}=c/\Omega$, inside which the magnetic field lines are usually considered to rotate rigidly with the magnetar.

If $r_\mathrm{m}<r_\mathrm{c}<r_\mathrm{L}$, materials at inner edge of the disk revolve faster than the local magnetic field lines and tend to be funneled before falling onto the surface of the magnetar. Thus, the magnetar gain its angular momentum and subsequently the co-rotation radius decreases until $r_\mathrm{c}\sim r_\mathrm{m}$. Conversely if $r_\mathrm{c}<r_\mathrm{m}<<r_\mathrm{L}$, the slow-rotating inner disk is speeded up to a super-Keplerian velocity by the magnetar, which results in mass ejection from disk and sharp spin-down of the magnetar (propeller regime). The spin-down of the magnetar, in return, leads to an increase of $r_\mathrm{c}$. Consequently, such a magnetar-disk system tends to evolve towards $r_\mathrm{c}=r_\mathrm{m}$ if the spin evolution of the magnetar is dominated by the interaction with the accretion disk. When $r_\mathrm{c}$ equals $r_\mathrm{m}$, the accreting magnetar would reach an equilibrium spin period (e.g., \citealp{2011ApJ...736..108P})
\begin{equation}
P_\mathrm{eq}=2\pi(GM)^{-5/7}R^{18/7}B^{6/7}\dot{M}^{-3/7}.
\label{eq: Peq}
\end{equation}
The equilibrium spin period is independent of the initial spin period of magnetar but correlates with the magnetic field strength and mass inflow rate. With $M=1.4$ M$_\odot$ and $R=12$ km, therefore, the accretion rate can be estimated from the $B-P$ distribution for accreting magnetars. 

Assuming a constant accretion rate, the evolutionary timescale, before a magnetar reaches such a spin equilibrium, can be estimated by \citep{2018ApJ...857...95M}
\begin{equation}
t_\mathrm{ev}\approx\frac{2\pi I/P_\mathrm{eq}}{\dot{M}(GMr_\mathrm{m})^{1/2}}=(GM)^{2/7}IR^{-8/7}B^{-8/7}\dot{M}^{-3/7},
\label{eq: Tev}
\end{equation}
where $I=0.35MR^2$ is the moment of inertia. The larger accretion rate and stronger magnetic field correspond to shorter timescale $t_\mathrm{ev}$. For an accreting magnetar that spins with an equilibrium period, an estimate of $t_\mathrm{ev}$ can be considered as the lower limit for the accretion timescale ($t_\mathrm{acc}$).

\section{Likelihood function for a linear fit}
\label{Sec: fit}

For fitting a linear function ($y=mx+c$) to data with errors on both variables, the likelihood function can be given by \citep{2005physics..11182D}
\begin{equation}
f\propto \prod \limits_{i} \frac{1}{\sqrt{\sigma^2_{y_i}+m^2\sigma^2_{x_i}}} \exp \left[-\frac{(y_i-mx_i-c)^2}{2(\sigma^2_{y_i}+m^2\sigma^2_{x_i})}\right],
\label{eq: Fit1}
\end{equation}
where $x_i$ and $y_i$ are the observational quantities, and $\sigma_{x_i}$ and $\sigma_{y_i}$ are the corresponding errors.

If there exists the extra variability of data, which can be parameterized as $\sigma_{v}$, the likelihood function is modified as \citep{2005physics..11182D}
\begin{equation}
f\propto \prod \limits_{i} \frac{1}{\sqrt{\sigma^2_{v}+\sigma^2_{y_i}+m^2\sigma^2_{x_i}}} \exp \left[-\frac{(y_i-mx_i-c)^2}{2(\sigma^2_{v}+\sigma^2_{y_i}+m^2\sigma^2_{x_i})}\right],
\label{eq: Fit2}
\end{equation}

By fitting the $\log B-\log P_0$ distribution of LGRB X-ray plateaus with Equation~(\ref{eq: Fit1}) as the likelihood function, we obtain the correlations, i.e. Equation~(\ref{eq: LGRB_iso}) for the isotropic wind model and Equation~(\ref{eq: LGRB_col}) for the collimated wind model. However, the extra variability of data ($\sigma_{v}$) should be considered for our SN sample, given the different models (see references listed in Table~\ref{Tab: sample} for details) and the lack of stringent constraint on the magnetar or $^{56}$Ni contribution. In fitting, the mean uncertainties of data is adopted as the errors for the likelihood function. For the two SNe~Ic, the uncertainties of $B$ and $P_0$ are not given in the literature, so we set the half of the values as the corresponding uncertainties. The fitting result for the SN sample is given in Equation~(\ref{Eq: BP_SNe}) with $\sigma_{v}=0.52^{+0.05}_{-0.04}$.

%\end{appendix}
\end{document}